\documentstyle[preprint,revtex,eqsecnum]{aps}
\begin{document}
\draft
\begin{title}
{}From wormhole to time machine:\\
Comments on Hawking's Chronology Protection Conjecture
\end{title}
\author{Matt Visser\cite{e-mail}}
\begin{instit}
Physics Department, Washington University, St. Louis,
Missouri 63130-4899
\end{instit}

\receipt{26 February 1992; Revised September 1992}

\begin{abstract}

The recent interest in ``time machines'' has been largely fueled
by the {\it apparent} ease with which such systems may be formed
in general relativity, given relatively benign initial conditions
such as the existence of traversable wormholes or of infinite cosmic
strings. This rather disturbing state of affairs has led Hawking
to formulate his {\it Chronology Protection Conjecture}, whereby
the formation of ``time machines'' is forbidden. This paper will
use several simple examples to argue that the universe appears to
exhibit a ``defense in depth'' strategy in this regard. For
appropriate parameter regimes Casimir effects, wormhole disruption
effects, and gravitational back reaction effects all contribute to
the fight against time travel.  Particular attention is paid to
the role of the quantum gravity cutoff. For the class of model
problems considered it is shown that the gravitational back reaction
becomes large before the Planck scale quantum gravity cutoff is
reached, thus supporting Hawking's conjecture.

\end{abstract}

\pacs{04.20.-q, 04.20.Cv, 04.60.+n; hepth@xxx/9202090 }
\narrowtext
\newpage
\section{INTRODUCTION}

This paper addresses Hawking's Chronology Protection Conjecture
\cite{HawkingI,HawkingII}, and some of the recent controversy
surrounding this conjecture \cite{Kim-Thorne}.

The recent explosion of interest in ``time machines'' is predicated
on the fact that it {\it appears} to be relatively easy to construct
such objects in general relativity. Time machines have been based
on the traversable wormholes of Morris and Thorne \cite{Morris-Thorne,MTY},
and on Gott's cosmic string construction \cite{Gott}.

These results have lead to a flurry of papers aimed at either making
one's peace with the notion of time travel
\cite{the-consortium,EKT,Friedman-Morris}, or of arguing that ---
despite {\it prima facie} indications --- the construction of time
machines is in fact impossible. Papers specifically addressing
Gott's cosmic string construction include
\cite{Carrol-Farhi-Guth,Deser-Jackiw-tHooft}. In a more general
vein Hawking's Chronology Protection Conjecture \cite{HawkingI,HawkingII}
is applicable to both wormhole constructions and cosmic string
constructions.

This paper will focus on Hawking's Chronology Protection Conjecture
specifically as applied to traversable wormholes. The discussion
will be formulated in terms of a simple model for a traversable
wormhole which enables one to carefully explain the {\it seemingly}
trivial manipulations required to turn a traversable wormhole into
a time machine. To help illustrate the manner in which nature might
enforce the Chronology Protection Conjecture the utility of two
particular approximation techniques will be argued. Firstly, it is
extremely useful to replace the original dynamical system with a
quasistationary approximation where one considers adiabatic variation
of parameters describing a stationary spacetime. It shall be argued
that this approximation captures essential elements of the physics
of time machine construction while greatly simplifying technical
computations, pedagogical issues, and conceptual issues. Secondly,
for a particular class of wormholes, once one is sufficiently close
to forming a time machine, it shall be shown that it is a good
approximation to replace the wormhole mouths with a pair of planes
tangent to the surface of the mouths. This approximation now reduces
all computations to variations on the theme of the Casimir effect
and permits both simple and explicit calculations in a well controlled
parameter regime.

The reader's attention is particularly drawn to the existence and
behaviour of a one--parameter family of closed ``almost'' geodesics
which thread the wormhole throat. The wormhole is deemed to initially
be well behaved so that the closed geodesics belonging to this
one--parameter family are initially spacelike. The putative
construction of a time machine will be seen to involve the invariant
length of the members of this family of geodesics shrinking to zero
--- so that the family of geodesics, originally spacelike, becomes
null, and then ultimately becomes timelike. Not too surprisingly,
it is the behaviour of the geometry as the length of members of
this family of closed geodesics shrinks to invariant length zero
that is critical to the analysis.

It should be emphasised that the universe appears to exhibit a
``defense in depth'' strategy in this regard. For appropriate ranges
of parameters describing the wormhole (such as masses, relative
velocities, distances, and time shifts) Casimir effects (geometry
induced vacuum polarization effects), wormhole disruption effects,
and gravitational back reaction effects all contribute to the fight
against time travel.

The overall strategy is as follows: start with a classical background
geometry on which some quantum fields  propagate --- this is just
the semi--classical quantum gravity approximation. In the vicinity
of the aforementioned family of closed spacelike geodesics the
vacuum expectation value of the renormalized stress--energy tensor
may be calculated approximately --- this calculation is a minor
variant of the standard Casimir effect calculation. The associated
Casimir energy will diverge as one gets close to forming a time
machine.  Furthermore, as the length of the closed spacelike
geodesics tends to zero the vacuum expectation value of the
renormalized stress--energy tensor itself diverges --- until
ultimately vacuum polarization effects become larger than the exotic
stress--energy required to keep the wormhole throat open ---
hopefully disrupting the wormhole before a time machine has a chance
to form.  Finally, should the wormhole somehow survive these
disruption effects, by considering linearized gravitational
fluctuations around the original classical background the gravitational
back reaction generated by the quantum matter fields can be estimated.
For the particular class of wormholes considered in this paper this
back reaction will be shown to become large before one enters the
Planck regime. It is this back reaction that will be relied upon
as the last line of defence against the formation of a time machine.
In this matter I am in agreement with Hawking \cite{HawkingI,HawkingII}.

\underbar{Units:} Adopt units where $c\equiv 1$, but all other
quantities retain their usual dimensionalities, so that in particular
$G=\hbar/m_P^2 = \ell_P^2/\hbar$.

\newpage
\section{FROM WORMHOLE TO TIME MACHINE}

The seminal work of Morris and Thorne on traversable wormholes
\cite{Morris-Thorne} immediately led to the realization that, given
a traversable wormhole, it {\it appeared} to be very easy to build
a time machine \cite{Morris-Thorne,MTY}. Indeed it so easy is the
construction that it {\it seemed} that the creation of a time
machine might be the generic fate of a traversable wormhole
\cite{Frolov-Novikov}.

In this regard it is useful to perform a {\sl gedankenexperiment} that
clearly and cleanly separates the various steps involved in
constructing a time machine:\\
--- step 0: acquire a traversable wormhole,\\
--- step 1: induce a ``time shift'' between the two mouths of the
wormhole,\\
--- step 2: bring the wormhole mouths close together.\\
It is only the induction of the ``time shift'' in step 1 that
requires intrinsically relativistic effects. It may perhaps be
surprising to realise that the apparent creation of the time machine
in step 2 can take place arbitrarily slowly in a non--relativistic
manner. (This observation is of course the ultimate underpinning
for one's adoption of quasistationary adiabatic techniques.)

\subsection{  Step 0: Acquire a traversable wormhole}

It is well known that topological constraints prevent the classical
construction of wormholes {\it ex nihilo} \cite{Geroch,Tipler76,Tipler77}.
The situation once one allows quantum gravitational processes is
less clear cut, but it is still quite possible that topological
selection rules might constrain
\cite{Gibbons-Hawking,Giulini,HorowitzI,HorowitzII} or even forbid
\cite{Visser90} the quantum construction of wormholes {\it ex
nihilo}. If such proves to be the case then even an arbitrarily
advanced civilization would be reduced to the mining of primordial
traversable wormholes --- assuming such primordial wormholes to
have been built into  the multiverse by the ``first cause''.
Nevertheless, to get the discussion started, assume (following
Morris and Thorne) that an arbitrarily advanced civilization has
by whatever means necessary acquired and continues to maintain a
traversable wormhole.

In the spirit of references \cite{Visser90,Visser89a,Visser89b} it is
sufficient, for the purposes of this chapter, to take an extremely
simple model for such a traversable wormhole --- indeed to
consider the wormhole as being embedded in flat Minkowski space
and to take the radius of the wormhole throat to be zero. Thus our
wormhole may be mathematically modelled by Minkowski space with
two timelike world lines identified.  For simplicity one may further
assume that initially the two mouths of the wormhole are at rest
with respect to each other, and further that the wormhole mouths
connect ``equal times'' as viewed from the mouth's rest frame.
Mathematically this means that one is considering (3+1) Minkowski
space with the following two lines identified:
\begin{equation}
\ell_1^\mu(\tau) = \bar\ell_0^\mu + {1\over2}\delta^\mu + V^\mu \tau,\\
\ell_2^\mu(\tau) = \bar\ell_0^\mu - {1\over2}\delta^\mu + V^\mu \tau,
\end{equation}
here $V^\mu$ is an arbitrary timelike vector, $\delta^\mu$ is
perpendicular to $V^\mu$ and so is spacelike, and $\bar\ell_0^\mu$
is completely arbitrary. The center of mass of the pair of wormhole
mouths follows the line
\begin{equation}
\bar\ell^\mu(\tau) = \bar\ell_0^\mu  + V^\mu \tau,
\end{equation}
and the separation of the wormhole mouths is described by the vector
$\delta^\mu$.

The present construction of course does not describe a time machine,
but one shall soon see that very simple manipulations {\it appear}
to be able to take this ``safest of all possible wormholes'' and
turn it into a time machine.

\subsection{  Step 1: Induction of a time--shift}

As the penultimate step in one's construction of a time machine it
is necessary to induce a (possibly small) ``time shift'' between
the two mouths of the wormhole. For clarity, work in the rest frame of
the wormhole mouths. After a suitable translation and Lorentz
transformation the simple model wormhole of step 0 can, without
loss of generality, be described by the identification $(t,0,0,0)
\equiv (t,0,0,\delta)$.  One wishes to change this state of affairs
to obtain a wormhole of type $(t,0,0,0) \equiv (t+T,0,0,\ell)$,
where $T$ is the time--shift and $\ell$ is the distance between
the wormhole mouths. Mathematically one wishes to force the vectors
$V^\mu$ and $\delta^\mu$ to no longer be perpendicular to each
other so that it is possible to define the time--shift to be $T = V^\mu
\delta_\mu$, while the distance between the mouths is $\ell = \Vert
(\delta^\mu{}_\nu + V^\mu V_\nu) \delta^\nu \Vert$.  Physically
this may be accomplished in a number  of different ways:

\subsubsection{The relativity of simultaneity}

Apply identical forces to the two wormhole mouths so that they
suffer identical accelerations (as seen by the initial rest frame).
It is clear that if the mouths initially connect equal times (as
seen by the initial rest frame), and if their accelerations are
equal (as seen by the initial rest frame), then the paths followed
by the wormhole mouths will be identical up to a fixed translation
by the vector $\delta^\mu$. Consequently the wormhole mouths will always
connect equal times --- as seen by the initial rest frame.
Mathematically this describes a situation where $\delta^\mu$ is a
constant of the motion while $V^\mu$ is changing. Of course, after the
applied external forces are switched off the two mouths of the
wormhole have identical four--velocities $V_f^\mu$ not equal to
their  initial four--velocities $V_i^\mu$. And by the relativity
of simultaneity equal times as viewed by the initial rest frame is
not the same as equal times as viewed by the final rest frame.  If
one takes the center of motion of the wormhole mouths to have been
accelerated from four--velocity $(1,0,0,0)$ to four--velocity
$(\gamma,0,0,\gamma\beta)$ then the relativity of simultaneity
induces a time--shift
\begin{equation}
T = \gamma \beta \delta,
\end{equation}
while (as seen in the final rest frame) the distance between the
wormhole mouths becomes $\ell = \gamma \delta$.

\subsubsection{Special relativistic time dilation}

Another way of inducing a time shift in the simple model wormhole
of step 0 is to simply move one of the wormhole mouths around and
to rely on special relativistic time dilation. This can either be
done by using rectilinear motion \cite{MTY} or by moving one of
the mouths around in a large circle \cite{Grischuk}. Assuming for
simplicity that one mouth is un-accelerated, that the other mouth
is finally returned to its initial four--velocity, and that the
final distance between wormhole mouths is the same as the initial
distance, then the induced time--shift is simply given by:
\begin{equation}
T = \int_i^f  (\gamma - 1) dt.
\end{equation}
Mathematically, $V^\mu$ and $\ell$ are unaltered, while $T$ and
hence $\delta^\mu$ are changed by this mechanism.

\subsubsection{General relativistic time dilation}

As an alternative to relying on special relativistic time dilation
effects, one may instead rely on the general relativistic time
dilation engendered by the gravitational redshift \cite{Frolov-Novikov}.
One merely places one of the wormhole mouths in a gravitational
potential for a suitable amount of time to induce a time--shift:
$$
T = \int_i^f  (\sqrt{g_{00}} - 1) dt.
$$

\subsubsection{Comment}

Of course, this procedure has not yet built a time machine. The
discussion so far has merely established that given a traversable
wormhole it is trivial to arrange creation of a traversable wormhole
that exhibits a time--shift upon passing through the throat ---
this time--shift can be created by a number of different mechanisms
so that the ability to produce such a time--shift is a very robust
result.

\subsection{  Step 2: Bring the wormhole mouths close together}

Having constructed a traversable wormhole with time--shift the
final stage of time machine construction is deceptively simple:
merely push the two wormhole mouths towards one another (this may
be done as slowly as is desired). A time machine forms once the
physical distance between the wormhole mouths $\ell$ is less than
the time--shift $T$. Once this occurs, it is clear that closed
timelike geodesics have formed --- merely consider the closed
geodesic connecting the two wormhole mouths and threading the
wormhole throat. As an alternative to moving the mouths of the
wormhole closer together one could of course arrange for the
time--shift to continue growing while keeping the distance between
the mouths fixed --- such an approach would obscure the distinctions
between steps 1 and 2.

The advantage of clearly separating steps 1 and 2 in one's mind is
that it is now clear that the wormhole mouths may be brought together
arbitrarily slowly --- in fact adiabatically --- and still {\sl
appear} to form a time machine. If one makes the approach adiabatic
(so that one can safely take the time--shift to be a constant of
the motion) then the wormhole may be mathematically modeled  by
Minkowski space with the lines $(t,0,0,z(t))$ and $(t+T,0,0,\ell_0-z(t))$
identified. The physical distance between the wormhole mouths is
$\ell(t)=\ell_0 - 2 z(t)$. The family of closed ``almost'' geodesics alluded
to in the introduction is just the set of geodesics (straight
lines), parameterized by $t$, connecting $\ell_1{}^\mu(t)$ with
$\ell_2{}^\mu(t)$. Specifically, with $\sigma\in[0,1]$:
\begin{eqnarray}
X^\mu(t,\sigma)
  =&& \sigma \ell_1{}^\mu(t) + \{1-\sigma\} \ell_2{}^\mu(t)\nonumber\\
  =&& \bigg(t + \{1-\sigma\} T, 0, 0,
            \{1-\sigma\} \ell_0 - \{1-2\sigma\}z(t)\bigg).
\end{eqnarray}
Here $\sigma$ is a parameter along the closed geodesics, while $t$
parameterizes the particular closed geodesic under consideration.
These curves are true geodesics everywhere except at
$(\sigma=0)\equiv(\sigma=1)$, the location of the wormhole mouths,
where there is a ``kink'' in the tangent vector induced by the
relative motion of the wormhole mouths. The invariant length of
the members of this family of closed geodesics is easily seen to
be:
\begin{equation}
\delta(t) = \Vert \ell_1{}^\mu(t) - \ell_2{}^\mu(t) \Vert
          = \Vert (T,0,0,\ell(t)) \Vert
	  = \sqrt{\ell(t)^2 - T^2}.
\end{equation}

Once $\ell(t) < T$  a time machine has formed. Prior to time machine
formation the vector $(T,0,0,\ell(t))$ is, by hypothesis, spacelike.
Therefore it is possible to find a Lorentz transformation to bring
it into the form $(0,0,0,\delta)$ with $\delta = \sqrt{\ell^2-T^2}$.
A brief calculation shows that this will be accomplished by a
Lorentz transformation of velocity $\beta = T/\ell(t) =
T/\sqrt{T^2+\delta^2(t)}$, so that $\gamma = \ell/\delta$ and
$\gamma\beta = T/\delta$.  In this new Lorentz frame the wormhole
connects ``equal times'', and this frame can be referred to as the
frame of simultaneity (synchronous frame). Note in particular, that
as one gets close to building a time machine $\delta(t)\to0$, that
$\beta\to1$, so that the velocity of the frame of simultaneity
approaches the speed of light.

\subsection{Discussion}

The construction process described above clearly separates the
different effects at work in the putative construction of time
machines. Given a traversable wormhole arbitrarily small special
and/or general relativistic effects can be used to generate a
time--shift.  Given an arbitrarily small time shift through a
traversable wormhole, arbitrarily slow adiabatic motion of the
wormhole mouths towards each other {\it appears} to be sufficient
to construct a time machine. This immediately leads to all of the
standard paradoxes associated with time travel and is very disturbing
for the state of physics as a whole.

As a matter of logical necessity precisely {\it one} of the following
alternatives must hold:\\
1: the Boring Physics Conjecture,\\
2: the Hawking Chronology Protection Conjecture,\\
3: the Novikov Consistency Conspiracy,\\
4: the Radical Rewrite Conjecture.

The Boring Physics Conjecture may roughly be formulated as: ``Suffer
not traversable wormholes to exist''.  Merely by asserting the
nonexistence of traversable wormholes  all time travel problems of
the particular type described in this paper go quietly away.
However, to completely forbid time travel requires additional
assumptions --  such as the nonexistence of (infinite length) cosmic
strings and limitations on the tipping over of light cones. For
instance, requiring the triviality of the fundamental group
$\pi_1({\cal M})$ (the first homotopy group) implies that the
manifold possesses no closed noncontractible loops.  Such a restriction
would preclude both (1) traversable wormholes (more precisely:
that class of traversable wormholes that connect a universe to
itself), and (2) cosmic strings (both finite and infinite). But
even this is not enough --- time travel can seemingly occur in
universes of trivial topology \cite{HawkingI,HawkingII,Godel}, so
even stronger constraints should be imposed.  In this regard
Penrose's version of the Strong Cosmic Censorship Conjecture
\cite{Penrose} implies the Causality Protection Conjecture via the
equivalence: (Strong Cosmic Censorship) $\Leftrightarrow$ (global
hyperbolicity) $\Rightarrow$ (strong causality) $\Leftrightarrow$
($\exists$ global time function). The ``Boring Physics Conjecture''
might thus best be formulated as Strong Cosmic Censorship together
with triviality of the fundamental group $\pi_1({\cal M})$. There
is certainly no experimental evidence against the Boring Physics
Conjecture, but this is a relatively uninteresting possibility. In
particular, considering the relatively benign conditions on the
stress--energy tensor required to support a traversable wormhole
it seems to be overkill to dispose of the possibility of wormholes
merely to avoid problems with time travel.

Less restrictively the Hawking Chronology Protection Conjecture
allows the existence of wormholes but forbids the existence of time
travel \cite{HawkingI,HawkingII}: ``Suffer not closed non-spacelike
curves to exist''.  If the Chronology Protection Conjecture is to
hold then there must be something wrong with the {\it apparently}
simple manipulations described above that {\it appeared} to lead
to the formation of a time machine. In fact, that is the entire
thrust of this paper --- to get some handle on precisely what might
go wrong.  It proves to be the case that the problem does not lie
with the induction of a time--shift --- as previously mentioned
this is a robust result that can be obtained through a number of
different physical mechanisms. Rather, it is the apparently very
simple notion of pushing the wormhole mouths together that is at
fault. A number of physical effects seem to conspire to prevent
one from actually bringing the wormhole mouths close together. This
will be discussed in detail in subsequent chapters.

More radically, the Novikov Consistency Conspiracy is willing to
countenance the existence of both traversable wormholes and time
travel but but asserts that the multiverse {\it must} be be consistent
no matter what --- this point of view is explored in
\cite{the-consortium,EKT,Friedman-Morris}: ``Suffer not an
inconsistency to exist''.

More disturbingly, once one has opened Pandora's box by permitting
closed timelike curves (time travel) I would personally be rather
surprised if something as relatively mild as the Novikov Consistency
Conspiracy would be enough to patch things up. Certainly the
sometimes expressed viewpoint that the Novikov Consistency Conspiracy
is the unique answer to the causality paradoxes is rather naive at
best.  The Novikov Consistency Conspiracy is after all still firmly
wedded to the notion of spacetime as a four--dimensional Hausdorff
differentiable manifold.

For a rather more violently radical point of view permit me to
propound the Radical Rewrite Conjecture wherein one posits a radical
rewriting of all of known physics from the ground up.  Suppose,
for instance, that one models spacetime by a non-Hausdorff
differentiable manifold. What does this mean physically? A
non-Hausdorff manifold has the bizarre property that the dimensionality
of the manifold is {\sl not} necessarily equal to the dimensionality
of the coordinate patches \cite{Chillingworth}.  From a physicist's
perspective, this idea has been explored somewhat by Penrose
\cite{Penrose}. Crudely put: while coordinate patches remain four
dimensional in such a spacetime, the dimensionality of the underlying
manifold is arbitrarily large, and possibly infinite. Local physics
remains tied to nicely behaved four dimensional coordinate patches.
Thus one can, for instance, impose the Einstein field equations in
the usual manner. Every now and then, however, a passing wave front
(generated by a ``branching event'') passes by and suddenly duplicates
the whole universe. It is even conceivable that a branching
non-Hausdorff spacetime of this type might be connected with or
connectable to the Everett  ``many worlds'' interpretation of
quantum mechanics \cite{Everett,Many-worlds}.

If one wishes an even more bizarre model of reality, one could
question the naive notion that the ``present'' has a unique fixed
``past history''. After all, merely by adding a time reversed
``branching event'' to our non-Hausdorff spacetime one obtains a
``merging event'' where two universes merge into one. Not only is
predicibility more than somewhat dubious in such a universe, but
one appears to have lost retrodictability as well. Even moreso than
time travel, such a cognitive framework would render the universe
unsafe for historians, as it would undermine the very notion of
the existence of a  unique ``history'' for the historians to
describe!

Such radical speculations might further be bolstered by the
observation that if one takes Feynman's ``sum over paths'' notion
of quantum mechanics seriously then all possible past histories of
the universe should contribute to the present ``state'' of the
universe.

I raise these issues, not because I particularly believe that that
is the way the universe works, but rather, because once one has
opened Pandora's box by permitting time travel, I see no particular
reason to believe that the only damage done to our notions of
reality would be something as facile as the Novikov Consistency
Conspiracy.

\newpage
\section{QUANTUM EFFECTS}
\subsection{Vacuum polarization effects}

Attention  is now drawn to the effect that the wormhole geometry
has on the propagation of quantum fields. Generically, one knows
that a non--trivial geometry modifies the vacuum expectation value
of the renormalized stress--energy tensor. One way of proceeding
is to recall that the wormhole is modelled by the identification
\begin{equation}
(t,0,0,0) \equiv (t+T,0,0,\ell).
\end{equation}
This formulation implies that one is working in the rest frame of
the wormhole mouths. For definiteness, suppose one is considering
a quantized scalar field propagating in this spacetime. One could
then, in principle, investigate solutions of d'Alembert's equation
in Minkowski space subject to these identification constraints,
find the eigenfunctions and eigenvalues, quantize the field and
normal order, and in this manner eventually calculate $<0|T^{\mu\nu}|0>$.
Alternatively, one could adopt point--splitting techniques. Such
calculations are decidedly non--trivial. For the sake of tractability
it is very useful to adopt a particular approximation that has the
effect of reducing the problem to a generalized Casimir effect
calculation.

Perform a Lorentz transformation of velocity $\beta = T/\ell$, so
that in this new Lorentz frame the wormhole connects ``equal times''.
In this frame of simultaneity (synchronous frame) the wormhole is
described by the identification
\begin{equation}
(\gamma t,0,0,\gamma\beta t) \equiv (\gamma t,0,0,\gamma\beta t + \delta).
\end{equation}

The discussion up to this point has assumed ``point like'' mouths
for the wormhole throat.  To proceed further one will need to take
the wormhole mouths to have some finite size, and will need to
specify the precise manner in which points on the wormhole mouths
are to be identified. Two particularly simple types of pointwise
identification are of immediate interest.

\underbar{(1) Synchronous identification:}\\ $\forall x_1 \in
\partial\Omega_1, x_2 \in \partial\Omega_2, x_1^\mu \equiv x_2^\mu
= x_1^\mu + s {\hat\delta}^\mu$, where $\hat\delta^\mu$ is a fixed
(spacelike) unit vector and $s$ is a parameter to be determined.
A particular virtue of synchronous identification is that once one
goes to the synchronous frame, all points on the wormhole mouths
are identified at ``equal times'', that is: $x_1 = (t,\vec x_1)
\equiv (t, \vec x_2) = (t, \vec x_1 + s \hat z) = x_2$.

\underbar{(2) Time-shift identification:}\\ $\forall x_1 \in
\partial\Omega_1, x_2 \in \partial\Omega_2, x_1 = (t, \vec x_1)
\equiv (t+T, \vec x_2) = x_2$, with $\vec x_1$, $\vec x_2$ ranging
over the mouths of the wormhole in a suitable manner,
This ``time--shift'' identification is the procedure adopted by
Kim and Thorne and is responsible for many of the technical
differences between that paper and this. A particularly unpleasant
side effect of ``time--shift'' identification is that there is no
one unique synchronous frame for the entire wormhole mouth.

Though these identification schemes apply to arbitrarily shaped
wormhole mouths, one may for simplicity, take the wormhole mouths
to be spherically symmetric of radius $R$ as seen in their rest
frames. The resulting spacetime is known as a {\sl capon} spacetime.
Adopting ``synchronous identification'', in the frame of simultaneity
the wormhole mouths are oblate spheroids of semi--major axis $R$,
and semi--minor axis $R/\gamma = R\delta/\ell$.  Let us work in
the parameter regime $\delta << \ell$ and $\delta << R$, then one
may safely approximate the wormhole mouths by flat planes.  One is
then seen to be working with a minor generalization of the ordinary
Casimir effect geometry --- a Casimir effect with a time--shift.
(For a nice survey article on the Casimir effect see
\cite{Centenary}.  See also the textbooks \cite{Birrell-Davies,Fulling}.)

Alternatively, one could work with the cubical wormholes of reference
\cite{Visser89a}.  When two of the flat faces of such a wormhole
are directly facing each other, the technical differences between
synchronous and time shift identifications vanish for those faces.
With this understanding the following comments can also be applied
to cubical wormholes with either identification scheme.

The model for the wormhole spacetime is now Minkowski space with
two planes identified:
\begin{equation}
(\gamma t,x,y,\gamma\beta t) \equiv (\gamma t,x,y,\gamma\beta t + \delta),
\label{sync}
\end{equation}
or, going back to the rest frame
\begin{equation}
(t,x,y,0) \equiv (t+T,x,y,\ell).
\label{rest}
\end{equation}
The net effect of (1) allowing for a finite size for the wormhole
throat, of (2) either adopting synchronous identification or of
restricting attention to flat faced wormholes, combined with (3)
the approximation $\delta <<R$, has thus been to replace the
identification of world--lines with the identification of planes.
The same effect could have been obtained by staying in the rest
frame, giving the wormhole mouths finite radius $R$, displacing
the wormhole mouths to $z=-R$ and $z=\ell+R$ respectively, and
letting $R\to\infty$. The advantage of the argument as presented
in the frame of simultaneity is that it makes clear that for any
synchronously identified traversable wormhole close enough to
forming a time machine this approximation becomes arbitrarily good.

Continuing to work in the frame of simultaneity, the manifest
invariance of the boundary conditions under rotation and/or reflection
in the $xy$ plane restricts the stress--energy tensor to be of the
form
\begin{equation}
<0|T_{\mu\nu}|0> = \left[
\begin{array}{cccc}
T_{tt}&0     &0     &T_{tz}\\
0     &T_{xx}&0     &0     \\
0     &0     &T_{xx}&0     \\
T_{tz}&0     &0     &T_{zz}
\end{array}
\right]
\end{equation}
By considering the effect of the boundary conditions one may write
eigenmodes of the d'Alambertian in the separated form
\begin{equation}
\phi(t,x,y,z) = e^{-i\omega t} e^{i k_x x} e^{i k_y y} e^{i k_z z},
\end{equation}
subject to the constraint $\phi(t,x,y,z) = \phi(t,x,y,z+\delta)$.
Therefore the boundary condition on $k_z$ implies the classical
quantization  $k_z = \pm n 2\pi/\delta$, while $k_x$ and $k_y$ are
unquantized ({\it i.e.} arbitrary and continuous). The equation
of motion constrains $\omega$ to be
\begin{equation}
\omega = \sqrt{(n 2\pi/\delta)^2 + k_\perp^2}.
\end{equation}
Now, recalling that $<0|T_{tz}|0> \propto <0|\partial_t \phi
\partial_z \phi|0>$, and performing a mode sum over $n$, it is
clear that the positive values of $k_z$ exactly cancel the negative
values so that $<0|T_{tz}|0> = 0$. Furthermore, one may note that
the quantization conditions on $\omega$ and $k_z$ depend only on
$\delta$ --- not on $\beta$ or $\gamma$. Thus, without loss of
generality one may immediately apply the result for $\beta=0$ to
the case $\beta\neq0$ and obtain
\begin{equation}
<0|T_{\mu\nu}|0> =
   - {\hbar k\over\delta^4} \left( \eta_{\mu\nu} - 4 n_\mu n_\nu \right)_.
\end{equation}
Here $n_\mu = \delta_\mu/\delta$ is the tangent to the closed
spacelike geodesic threading the wormhole throat. In the frame of
simultaneity $n_\mu = (0,0,0,1)$, while in the rest frame
$n_\mu=(-T/\delta,0,0,\ell/\delta)$. The constant $k$ is a
dimensionless numerical factor that it is not worth the bother to
completely specify. In a more general context the argument given
above applies not only to scalar fields but also to fields of
arbitrary spin that are (approximately) conformally coupled. In
general $k$ will depend on the nature of the applied boundary
conditions (twisted or untwisted), the spin of the quantum field
theory under consideration, etc. Furthermore, as $\delta$ decreases,
massive particles may be considered as being approximately massless
once $\hbar/\delta >> m$. Thus $k$ should be thought of as changing
in stepwise fashion as $\delta\to0$. This behaviour is entirely
analogous to the behaviour of the $R$ parameter in $e^+\, e^-$
annihilation. A complete specification of $k$ would thus require
complete knowledge of the asymptotic behaviour of the spectrum of
elementary particles, and it is for this reason that one declines
further precision in the specification of $k$.

The form of the stress--energy tensor can also be determined by
noting that the identifications (\ref{sync}) and (\ref{rest})
describing the wormhole can be easily solved by considering an
otherwise free field in Minkowski space subject to the constraint
\begin{equation}
\phi(x^\mu) \equiv \phi(x^\mu \pm n\delta^\mu) \\ n=0,1,2,...
\end{equation}
Since these constraints do not depend on $V$, the four velocity of
the wormhole mouth, one may directly apply the results obtained for
the ordinary Casimir effect \cite{Centenary,Birrell-Davies,Fulling}.

At this stage it is also important to realise that the sign of $k$
is largely immaterial. If $k$ is positive then one observes a
positive Casimir energy which tends to repel the wormhole mouths,
and so tends prevent the formation of a time machine. On the other
hand if $k$ is negative one finds a large positive stress threading
the wormhole throat. This stress tends to act to collapse the
wormhole throat, and so also tends to prevent the formation of a
time machine.

Note the similarities --- and differences --- between this result
and those of Hawking \cite{HawkingI,HawkingII} and Kim and Thorne
\cite{Kim-Thorne}.   The calculation presented here is in some ways
perhaps more general in that it is clear from the preceding discussion
that this calculation is capable of applying to the generic class
of ``synchronously identified'' wormholes once $\delta << R$.
Furthermore, instead of the value of the stress--energy tensor
being related to the temporal distance to the ``would be Cauchy
horizon'', it is clear in this formalism that it is the existence
and length of closed spacelike geodesics that controls the magnitude
of the stress--energy tensor.

The limitations inherent in the type of wormhole model currently
under consideration should also be made clear: the model problem
is optimally designed to make it easy to thwart the formation of
a time machine (i.e. closed timelike curves).  It is optimal for
thwarting in two senses: (1) the choice of synchronous identification
prevents defocussing of light rays when they pass through the
wormhole, (i.e. it makes $f=0$ in the notation of Hawking
\cite{HawkingI,HawkingII}). If an advanced civilization were to
try to create a time machine, that civilization presumably would
optimize their task by choosing $f$ arbitrarily large, and not
optimize Nature's ability to thwart by choosing $f=0$.  (2) the
adiabatic approximation requires the relative motion of the wormhole
mouths to be arbitrarily slow, (i.e. $h$ arbitrarily close to zero,
in the notation of Hawking \cite{HawkingI,HawkingII}).  This is an
idealized limit;  and again, it is the case that this idealization makes it
easier for Nature to thwart the time machine formation, because
it gives the growing vacuum polarization an arbitrarily long time
to act back on the spacetime and distort it.

\subsection{$k$ positive --- repulsive Casimir energy}

For the synchronously identified model wormholes we have been
discussing one may calculate the four--momentum associated with
the stress--energy tensor by
\begin{equation}
P^\mu = \oint <0|T^{\mu\nu}|0> d\Sigma_\nu
      = <0|T^{\mu\nu}|0> \pi\; R^2\; \delta\; n^\mu_\perp.
\end{equation}
Here $n^\mu_\perp = (1,0,0,0)$ in the frame of simultaneity, so
that $n_\perp$ is perpendicular to $n$. Thus
\begin{equation}
P^\mu = {\hbar k \over\delta^4}\; \pi\; R^2\; \delta\; n^\mu_\perp.
\end{equation}
By going back to the rest frame one obtains
\begin{equation}
P^\mu = {\hbar k \over\delta^4} \pi R^2 \delta
\left({\ell\over\delta},0,0,{T\over\delta}\right),
\end{equation}
so that one may identify the Casimir energy as
\begin{equation}
E_{Casimir} = P^\mu V_\mu = {\hbar k \pi R^2 \ell \over \delta^4}.
\end{equation}

Take $k >0$ for the sake of discussion, then this Casimir energy
would  by itself give an infinitely repulsive hard core to the
interaction between the wormhole mouths. However one should exercise
some care. The calculation is certainly expected to break down for
$\delta < \ell_P$, so that one may safely conclude only that there
is a finite but large potential barrier to entering the full quantum
gravity regime.  Fortunately this barrier is in fact very high.
For $T>>T_P$ one can estimate
\begin{equation}
E_{barrier} \approx E_P \left({R\over \ell_P}\right)^2
                        \left({T\over T_P} \right),
\end{equation}
while for $T<<T_P$ ({\it i.e.} $T\approx 0$) one may estimate
\begin{equation}
E_{barrier} \approx E_P \left({R\over \ell_P}\right)^2.
\end{equation}

For macroscopic wormholes these barriers are truly enormous. For
microscopic wormholes ($R << \ell_P$) one really doesn't care what
happens. Firstly because it is not too clear wether wormholes with
$R << \ell_P$ can even exist. Secondly, because even if such
wormholes do exist, they would in no sense be traversable
\cite{Visser90}, and so would be irrelevant to the construction of
usable time machines.  The non--traversability of such microscopic
wormholes is easily seen from the fact that they could only be
probed by particles of Compton wavelength much less that their
radius, implying that
\begin{equation}
E_{probe} > \hbar/R >> E_P.
\end{equation}

On the other hand, for macroscopic wormholes one may safely assert that
\begin{equation}
R/\ell_p > 2m/m_P,
\end{equation}
since otherwise each wormhole mouth would be enclosed by its own
event horizon and the wormhole would no longer be traversable.
Combining these results, it is certainly safe to assert that the
barrier to full quantum gravity satisfies
\begin{equation}
E_{barrier} >>  E_P \left({2m\over m_P}\right)^2
               \approx {m^2\over m_P}.
\end{equation}

To get a better handle on the parameter regime in question, consider
the combined effects of gravity and the Casimir repulsion. In
linearized gravity the combined potential energy is
\begin{equation}
V = - \left({m\over m_P}\right)^2 {\hbar\over \ell}
    + {\hbar k \pi R^2 \ell \over (\ell^2-T^2)^2}.
\end{equation}
The Casimir energy dominates over the gravitational potential energy
once
\begin{equation}
\delta^4 \approx {k \pi R^2 \ell^2 \over (m/m_P)^2 }
       > k \pi \ell_P^2 \ell^2,
       \\ {\it i.e.} \\
\delta \ge \sqrt{\ell_P \ell}.
\end{equation}

One may safely conclude: (1) for $k$ positive there is an enormous
potential barrier to time machine formation, (2) for ``reasonable''
wormhole parameters the repulsive Casimir force dominates long
before one enters the Planck regime.

\subsection{$k$ negative --- wormhole disruption effects}

At first glance, should the constant $k$ happen to be negative,
one would appear to have a disaster on one's hands. In this case
the Casimir force seems to act to help rather than hinder time
machine formation. Fortunately yet another physical effect comes
into play. Recall, following Morris and Thorne \cite{Morris-Thorne}
that a traversable wormhole must be threaded by some exotic stress
energy to prevent the throat from collapsing. In particular, at
the throat itself (working in Schwarzschild coordinates) the
stress--energy tensor takes the form
\begin{equation}
T_{\mu\nu} = {\hbar\over \ell_P^2 R^2} \left[
\begin{array}{cccc}
\xi\quad&0\quad  &0\quad&0     \\
0       &\chi    &0     &0     \\
0       &0       &\chi  &0     \\
0       &0       &0     &-1
\end{array}
\right]
\end{equation}
On general grounds $\xi <1$, while $\chi$ is unconstrained.  On the
other hand, the vacuum polarization effects just considered contribute
to the stress--energy tensor in the region between the wormhole
mouths an amount
\begin{equation}
<0|T_{\mu\nu}|0> = {\hbar k \over \delta^4}\left[
\begin{array}{cccc}
4(T/\delta)^2+1\quad&0\quad&0\quad&4T\ell/\delta^2\\
0                   &-1    &0     &0             \\
0                   &0     &-1    &0             \\
4T\ell/\delta^2     &0     &0     &4(\ell/\delta)^2-1
\end{array}
\right]
\end{equation}
In particular, the tension in the wormhole throat, required to
prevent its collapse is $\tau=\hbar/(\ell_P^2 R^2)$, while the
tension contributed by vacuum polarization effects is
\begin{equation}
\tau = - {\hbar k \over \delta^4} \left(4(\ell/\delta)^2-1\right)
     \approx {- 4 \hbar k \ell^2 \over \delta^6}
\end{equation}
Vacuum polarization effects dominate over the wormhole's internal
structure once
\begin{equation}
\delta^3 < \ell_P  \ell R /\sqrt{|k|},
\end{equation}
and it is clear that this occurs well before reaching the Planck
regime.

By looking at the {\it sign} of the tension it appears that for
negative values of $k$ these vacuum polarization effects will tend
to decrease $R$, that is, will tend to collapse the wormhole throat.
Conversely, for positive values of $k$ these vacuum polarization
effects would appear to tend to make the wormhole grow.

Concentrate on negative values of $k$. Well before $\delta$ shrinks
down to the Planck length, the vacuum polarization will have severely
disrupted the internal structure of the wormhole --- presumably
leading to wormhole collapse. It should be pointed out that there
is no particular need for the collapse to proceed all the way down
to $R=0$ and subsequent topology change. It is quite sufficient
for the present discussion if the collapse were to halt at $R\approx
\ell_P$. In fact, there is evidence, based on minisuperspace
calculations \cite{Visser90a,Visser91a,Visser91b}, that this is
indeed what happens. If indeed collapse is halted at $R\approx
\ell_P$ by quantum gravity effects then the universe is still safe
for historians since there is no reasonable way to get a physical
probe through a Planck scale wormhole~\cite{Visser90}.

\subsection{Summary}

Vacuum polarization effects become large as $\delta\to0$. Depending
on an overall undetermined sign either (1) there is an arbitrarily
large force pushing the wormhole mouths apart, or (2) there are
wormhole disruption effects at play which presumably collapse the
wormhole throat down to the size of a Planck length. Either way,
usable time machines are avoided.

Unfortunately, because the class of wormholes currently under
consideration is optimally designed to make it easy to thwart the
formation of a time machine, it is unclear to what extent one may
draw generic conclusions from these arguments.

\newpage
\section{GRAVITATIONAL BACK REACTION}

There is yet another layer to the universe's ``defense in depth''
of global causality.  Consider linearized fluctuations around the
locally flat background metric describing the synchronously identified
model wormhole.
\begin{equation}
g_{\mu\nu} = \eta_{\mu\nu} + \bar h_{\mu\nu} - {1\over2} \bar h \eta_{\mu\nu}.
\end{equation}
By going to the transverse gauge $\bar h^{\mu\nu},_\nu=0$, one may
write the linearized Einstein field equations as
\begin{equation}
\Delta\bar h_{\mu\nu} = {-16\pi\ell_P^2 \over \hbar} <0|T_{\mu\nu}|0>.
\end{equation}
Now, working in the rest frame of the wormhole mouths, the time
translation invariance of the geometry implies $\partial_t \bar
h_{\mu\nu} = 0$. Similarly, within the confines of the Casimir
geometry approximation, the translation symmetry in the $x$ and
$y$ directions implies $\partial_x \bar h_{\mu\nu} = 0 = \partial_y
\bar h_{\mu\nu}$. Thus the linearized Einstein equations reduce to
\begin{equation}
\partial_z{}^2 \bar h_{\mu\nu} =  { 16\pi k \ell_P^2 \over \delta^4}
           \left(\eta_{\mu\nu} - 4 n_\mu n_\nu \right).
\end{equation}
Note that the tracelessness of $T_{\mu\nu}$ implies the tracelessness
of $\bar h_{\mu\nu}$ so that $\bar h_{\mu\nu} = h_{\mu\nu}$. Using
the boundary condition that $h_{\mu\nu}(z=0,t=0) = h_{\mu\nu}(z=\ell,t=T)$
the linearized Einstein field equations integrate to
\begin{equation}
 h_{\mu\nu}(z) = {16\pi\over2}(z-\ell/2)^2 \;{k \ell_P^2 \over \delta^4}
           \; \left(\eta_{\mu\nu} - 4 n_\mu n_\nu \right).
\end{equation}
To estimate the maximum perturbation of the metric calculate
\begin{equation}
 \delta g_{\mu\nu} = h_{\mu\nu}(z=0) =  h_{\mu\nu}(z=\ell) =
   {16\pi k \ell_P^2 \ell^2 \over 8\delta^4}
           \left(\eta_{\mu\nu} - 4 n_\mu n_\nu \right).
\end{equation}
In particular
\begin{equation}
 \delta g_{tt} =    {-16\pi k \ell_P^2 \ell^2 \over 8\delta^4}
           \left({1 + 4 {T^2\over\delta^2}} \right).
\end{equation}

Now if $T>>T_P$, which is the case of interest for discussing the
putative formation of time machines, this back reaction certainly
becomes large well before one enters the Planck regime, thus
indicating that the gravitational back reaction becomes large and
important long before one needs to consider full quantum gravity
effects.

Even if $T=0$ one has $\delta g_{tt} = -(16\pi/8) k (\ell_P^2 / \ell^2)$.
While the back reaction is in this case somewhat smaller, it still
becomes large near the Planck scale.

It might be objected that the frame dependent quantity $\delta
g_{tt}$ is not a good measure of the back reaction. {\it Au
contraire}, one need merely consider the manifestly invariant object
\begin{equation}
\varphi = h_{\mu\nu} V^\mu V^\nu.
\end{equation}
Here $V^\mu$ is the four velocity of the wormhole throat, and
$\varphi$ is the physical gravitational potential governing the
motion of the wormhole mouths. For instance, the geodesic acceleration
of the wormhole mouths is easily calculated to be
\begin{equation}
a^\mu = V^\mu{}_{;\nu} V^\nu = \Gamma^\mu_{\alpha\beta} V^\alpha V^\beta =
-{1\over2} \nabla^\mu \varphi.
\end{equation}

Because of the delicacy of these particular issues, it may be
worthwhile to belabor the point. Suppose one redoes the entire
computation in the frame of simultaneity. One still has
\begin{equation}
\Delta\bar h_{\mu\nu} = {-16\pi\ell_P^2 \over\hbar} <0|T_{\mu\nu}|0>.
\end{equation}
Where now the boundary conditions are $h_{\mu\nu}(z=\beta t,t) =
h_{\mu\nu}(z=\beta t + \delta,t)$.  This immediately implies that
$h_{\mu\nu}$ is a function of $(z-\beta t)$. Indeed the linearized
Einstein equations integrate to
\begin{equation}
 h_{\mu\nu}(z,t) =  {16\pi\over2}{(z-\beta t - \delta/2)^2 \over (1-\beta^2)}
                 \;{k \ell_P^2 \over \delta^4}
                 \;\left(\eta_{\mu\nu} - 4 n_\mu n_\nu \right).
\end{equation}
So that at the mouths of the wormhole, $(z=\beta t)\equiv(z=\beta
t +\delta)$, factors of $\delta$ and $\gamma = 1/\sqrt{1-\beta^2}$
combine to yield
\begin{equation}
 \delta g_{\mu\nu} = h_{\mu\nu}(z=\beta t,t)
                   =  h_{\mu\nu}(z=\beta t +\ell,t)
		   = {16\pi k \ell_P^2 \ell^2 \over 8\delta^4}
                    \left(\eta_{\mu\nu} - 4 n_\mu n_\nu \right).
\end{equation}
Which is the same tensor as the result calculated in the rest frame,
(as of course it must be).

The only real subtlety in this case is to realise that, as previously
argued, $\delta g_{tt}$ as measured in the frame of simultaneity
is {\it not} a useful measure of the gravitational back reaction.

Yet another way of seeing this is to recall that in linearized
Einstein gravity the gravitational potential of a pair of point
particles is determined by their masses, relative separation, and
their four velocities by \cite{Nieto-Goldman,Macrae-Riegert}
\begin{equation}
V = \left\{ {G m_1 m_2 \over r}  \right\}
    \left\{ {2(V_1 \cdot V_2)^2 -1 \over \gamma_1 \gamma_2} \right\}.
\end{equation}
Applied to the case presently under discussion, one immediately
infers that the difference between the synchronous frame and the
rest frame induces an enhancement factor of $\gamma^2$.

While the four velocity of the wormhole throat does not enter into
the computation of the stress energy tensor it is important to
realise that the four velocity of the throat does, via the naturally
imposed boundary conditions, have an important influence on the
gravitational  back reaction.

In any event, the gravitational back reaction will radically alter
the spacetime geometry long before a time machine has the chance
to form. Note that for $k$ negative, so that the Casimir energy is
negative and attractive, the {\sl sign} of $\delta g_{tt}$ obtained
from this linearized analysis indicates a repulsive back reaction
and furthermore hints at the formation of an event horizon should
$\delta$ become sufficiently small. Unfortunately a full nonlinear
analysis would be necessary to establish this with any certainty.

\newpage
\section{GENERALITIES}

To put the analysis of this paper more properly in perspective it
is useful to abstract the essential ingredients of the calculation.
Consider a stationary, not necessarily static, but otherwise
arbitrary Lorentzian spacetime of nontrivial topology. Specifically,
assume that $\pi_1({\cal M}) \neq {e}$, that is, that the first
homotopy group (the fundamental group) is nontrivial. By definition
the non--triviality of $\pi_1({\cal M})$ implies the existence of
closed paths not homotopic to the identity. This is the {\it sine
qua non} for the existence of a wormhole. By smoothness arguments
there also exist smooth closed paths not homotopic to the identity.
Take one of these smooth closed paths and extremize its length in
the Lorentzian metric.  One infers the existence of at least one
smooth closed geodesic in any spacetime with nontrivial fundamental
group. If any of these closed geodesics is timelike or null then
the spacetime is diseased and should be dropped from consideration.
Furthermore, since the metric can be expressed in a $t$ independent
manner one may immediately infer the existence of an infinite
one--parameter family of closed geodesics parameterized by $t$. A
slow adiabatic ({\it i.e.} quasi--stationary) variation of the
metric will preserve this one parameter family, though the members
of this family may now prove to be only approximately geodesics.
The formation of a time machine is then signalled by this one
parameter family of closed ``almost'' geodesics switching over from
spacelike character to null and then timelike character.

If one is desirous of proceeding beyond the adiabatic approximation
this can be done at the cost of additional technical machinery.
Consider now  a completely  arbitrary Lorentzian spacetime of
nontrivial topology. Pick an arbitrary base point $x$. Since
$\pi_1({\cal M})$ is nontrivial there certainly exist closed paths
not homotopic to the identity that begin and end at $x$. By smoothness
arguments there also exist smooth closed paths not homotopic to
the identity --- though now there is no guarantee that the tangent
vector is continuous as the path passes through the point $x$ where
it is pinned down.  Take one of these smooth closed paths and
extremize its length in the Lorentzian metric. One infers the
existence of many smooth (except at the point $x$) closed geodesics
passing through every point $x$ in any spacetime with nontrivial
fundamental group. Again, if any of these closed geodesics is
timelike or null then the spacetime is diseased and should be
dropped from consideration.

To get a suitable one parameter family of closed geodesics that
captures the essential elements of the geometry, suppose merely
that one can find a well defined throat for one's Lorentzian
wormhole. Place a clock in the middle of the throat. At each time
$t$ as measured by the wormhole's clock there exists a closed
``pinned'' geodesic threading the wormhole and closing back on
itself in ``normal space''.  This geodesic will be smooth everywhere
except possibly at the place that it is ``pinned'' down by the
clock. This construction thus provides one with a one parameter
family of closed geodesics suitable application of the preceding
analysis.

Pick one of these closed geodesics. Let $\sigma\in[0,1]$ be a
parameterization of the geodesic. Extend this to a coordinate patch
$\{t,x,y,\sigma\}$ covering a tube--like neighbourhood surrounding
the geodesic. By adopting the spacelike generalization of comoving
coordinates one may without loss of generality write the metric in
the form
\begin{equation}
ds^2 = g_{ij}(t,x,y,\sigma) dx^i dx^j + \delta^2(t,x,y)d\sigma^2.
\end{equation}
Here $dx^i \in \{dt,dx,dy\}$, while the fact that the curve $x=y=t=0$
is a geodesic is expressed by the statement $\partial_i\delta(0,0,0)=0$.
By considering the three dimensional Riemann tensor on slices of
constant $\sigma$ one may define a quantity
\begin{equation}
R^{-2} = \max_{\sigma\in[0,1]} \sqrt{R_{ijkl} R^{ijkl}}.
\end{equation}
This quantity $R$ estimates the minimum ``radius of curvature''
of the constant $\sigma$ hypersurfaces and so may usefully be
interpreted as a measure of the minimum radius of the wormhole
throat.

For a traversable wormhole, precursor to a traversable time machine,
one wishes $R$ to be macroscopic and to remain so. Now consider
what happens as $\delta$ shrinks. Eventually $\delta<<R$, at which
stage the length of the closed geodesic is much less than the length
scale of the transverse dimensions of the wormhole throat.  By
going to Riemann normal coordinates one may now approximate the
metric by
\begin{equation}
ds^2 = \eta_{ij} dx^i dx^j + \delta^2(t,x,y) d\sigma^2, \\ (x,y << R),
\end{equation}
which is just a generalization of the Casimir geometry expressed
in the synchronous frame. Nasty ``fringe'' effects occur for $x,y
\ge R$, but these may be quietly neglected as is usual and reasonable
within the confines of the Casimir approximation. Of greater
significance is the behaviour of $\delta(t,x,y)$ as a function of
$t$, $x$, and $y$. If the variation of $\delta(t,x,y)$ is not
particularly rapid one may safely make the further approximation
of modelling the situation by the ordinary Casimir effect. In that
case the conclusions of the previous chapter then follow in this
more general context.

This happy circumstance is automatically fulfilled if if the wormhole
mouths are synchronously identified, since then $\delta(t,x,y) \leq
\delta(t,0,0) + O(R/\gamma) = \delta(t,0,0) [1+O(R/\ell)]$. On the
other hand, the time shift identification adopted by Kim and Thorne
leads to a very rapid growth of $\delta(t,x,y)$. This rapid variation
of $\delta(t,x,y)$ with position makes analysis considerably more
difficult.  Fortunately, it is still relatively easy to convince
oneself that the vacuum expectation value of the renormalized stress
energy tensor along the central geodesic must be proportional to
$\delta^{-4}$. Therefore, at least along the central geodesic, the
exotic matter supporting the wormhole throat is overwhelmed by the
renormalized stress energy. The extent to which this disruption
along the central geodesic might lead to disruption of the
traversability of the wormhole is unfortunately not calculable by
the present techniques.

The whole point of this aspect of the discussion is, of course, to
see what exactly is special about the simple model wormholes
considered earlier --- the analysis presented in this paper is seen
to be limited to adiabatic (spatial and temporal) changes in
$\delta(t,x,y)$.

\newpage
\section{COMPARISONS WITH PREVIOUS WORK}

There are several manners in which one might attempt to parameterize
the strength of the divergences in the stress--energy tensor and
the gravitational back reaction. The estimates of Kim and Thorne
\cite{Kim-Thorne} make extensive use of the ``proper time to the
Cauchy horizon'' as measured by an observer who is stationary with
respect to one of the wormhole mouths.  Hawking \cite{HawkingI,HawkingII}
advocates the use of an ``invariant distance to the Cauchy horizon'',
which is equal to the proper time to the Cauchy horizon as measured
by an observer who is stationary with respect to the frame of
simultaneity. On the other hand, this paper has focussed extensively
on the invariant length of closed geodesics.

The calculations of this paper suggest that the use of the ``proper
time to the Cauchy horizon'', whether measured by an observer in
the rest frame or in the synchronous frame is not a useful way of
parameterizing the singularities in the stress--energy tensor. To
see this, transcribe the ``order of magnitude estimates'' of Kim
and Thorne \cite{Kim-Thorne} into the notation of this paper ($\Delta
t$ is the proper distance to the Cauchy horizon as measured in the
rest frame).  Consider a pair of slowly moving wormhole mouths with
relative velocity $\beta_{rel}$ (not to be confused with the totally
different $\beta \equiv T/\ell$ associated with the transformation
from the rest frame to the synchronous frame). For a small relative
velocity, the wormhole is modeled by the identification of world--lines
\begin{equation}
(t,0,0,0)\equiv(t+T,0,0,\ell-\beta_{rel}\, t),
\end{equation}
so that $\delta^2(t) = (\ell - \beta_{rel}\, t)^2 - T^2$. (There is
nothing particularly sacred about taking a small relative velocity
--- it just simplifies life in that one only has a simple linear
equation to solve instead of a quadratic.) The Cauchy horizon forms
when $\delta^2(\Delta t)=0$, that is, when
\begin{equation}
\Delta t = {(\ell - T) \over \beta_{rel}}.
\end{equation}
Resubstituting into $\delta^2\equiv\delta^2(t=0)$, noting that
$\ell+T\approx 2\ell$, and being careful to retain all factors of
$\beta_{rel}$, one obtains for the Kim-Thorne estimates
\begin{eqnarray}
\delta^2          &&\approx 2\ell (\ell - T)
                  = 2\ell\; \beta_{rel}\; \Delta t.\\
<0|T^{\mu\nu}|0>  &&\approx (R/\ell)^\zeta \;
                  {1\over\ell (\beta_{rel}\Delta t)^3}, \\
                  && =(R/\ell)^\zeta \;
		  {1\over \ell (\ell - T)^3},\\
\delta g_{\mu\nu} &&\approx (R/\ell)^\zeta \;
                  {\ell_P^2\over\ell\beta_{rel}\Delta t}, \\
                  &&= (R/\ell)^\zeta \;
		  {\ell_P^2\over\ell(\ell-T)}.
\end{eqnarray}
Here $\zeta$ is an integer exponent that depends on the homotopy
class of the closed geodesic, and on whether or not one is close
to either mouth of the wormhole. (The factors of $\beta_{rel}$ were
unfortunately omitted in the discussion portion of reference
\cite{Kim-Thorne}.)

To see the inadequacy of the use of $\Delta t$ at a conceptual
level, observe that the use of $\Delta t$ intrinsically ``begs the
question'' of the creation of a time machine by explicitly asserting
the existence of a Cauchy horizon and then proceeding to measure
distances from that presumed horizon. In particular, one may consider
a wormhole in which one keeps the distance between the mouths fixed
({\it i.e.} set the relative velocity of the mouths to zero). By
definition, this implies that the Cauchy horizon never forms and
that $\Delta t = +\infty$.  (If one really wishes to be technical,
consider a collection of wormholes of the type considered by Kim
and Thorne and take the limit as the relative velocity goes to
zero).

To get a better handle on the actual state of affairs, go to the
Casimir limit by adopting synchronous identification and taking
$\delta << R$. In this case one may safely neglect the geometrical
factors $(R/\ell)^\zeta$.   By adopting a point spitting regularization
the (renormalized) propagator may easily be seen to be approximated
by
\begin{equation}
<0|\phi(x)\phi(x)|0> \approx \delta^{-2}.
\end{equation}
The stress--energy tensor is then computable by taking two derivatives
of the propagator, yielding
\begin{equation}
<0|T^{\mu\nu}|0>  \approx \hbar\delta^{-4} t^{\mu\nu}.
\end{equation}
Here $t^{\mu\nu}$ is a dimensionless tensor built up out of the
metric and tangent vectors to the closed spacelike geodesic. While
the {\it components} of $t^{\mu\nu}$ may be large in some Lorentz
frames, the only sensible invariant measure of the size, $t^{\mu\nu}
t_{\mu\nu}$, is of order one. In particular, looking at the $tt$
component, as viewed from the rest frame of one of the wormhole
mouths, yields
\begin{equation}
t^{tt} \approx \ell^2 \delta^{-2} \\  \Rightarrow \\
<0|T^{tt}|0>  \approx \hbar\ell^2\delta^{-6}
              \approx \hbar \ell^{-1} (\ell - T)^{-3}.
\end{equation}
This is the analogue of the estimate of Kim and Thorne, (including
insertion of the appropriate factor of $\beta_{rel}$). However,
one sees that this estimate is somewhat misleading in that it is
a componentwise estimate that is highly frame dependent.

Returning to the tensorial formulation, a double spatial integration
gives an estimate for linearized metrical fluctuations
\begin{equation}
\delta g_{\mu\nu} \approx \ell_P^2 \ell^2 \delta^{-4} t^{\mu\nu}.
\end{equation}
This estimate (which is backed up by the earlier explicit calculation)
is radically different from that of Kim and Thorne. The difference
can be traced back to the choice of synchronous identification
versus time shift identification, and the effect that these different
identification procedures have on the van Vleck determinant.

Turning to other matters: To understand Hawking's ``invariant''
distance to the Cauchy horizon, go to the synchronous frame. One
requires $\beta_{rel}^{synch} << \beta\equiv T/\ell$, so that the
wormhole may be described by the identification
\begin{equation}
\bigg(t,0,0,\beta t\bigg) \equiv
\bigg(t,0,0, (\beta - \beta_{rel}^{synch}) t + \delta\bigg),
\end{equation}
then $\delta(t)^2 = (\delta - \beta_{rel}^{synch} t)^2$. (Warning:
$\beta_{rel}^{synch}$ is now the relative velocity as measured in
the synchronous frame.) The Cauchy horizon forms at
\begin{equation}
t= \delta/\beta_{rel}^{synch}.
\end{equation}
Hawking now takes this object $t=\delta/\beta_{rel}^{synch}$, which
is in fact an invariant measure of the distance to the Cauchy
horizon, as his parameter governing the strength of the singularities
encountered when trying to build a time machine. Of course, this
parameter suffers from deficiencies analogous to the Kim--Thorne
parameter in that it is intrinsically incapable of properly reflecting
the divergence structure that is known to occur in the simple
stationary (or indeed quasistationary) models considered in this
paper.

Turning to the question of the quantum gravity cutoff, Kim and
Thorne assert that this cutoff occurs at
\begin{equation}
\Delta t = {(\ell-T)\over\beta_{rel}}\approx\ell_P \\ \Rightarrow \\
\ell - T \approx \beta_{rel}\; \ell_P.
\end{equation}
Hawking claims that the cutoff occurs at
\begin{equation}
t= \delta/\beta_{rel}^{synch} \approx \ell_P \\ \Rightarrow \\
\delta \approx \beta_{rel}^{synch}\; \ell_P.
\end{equation}
I beg to differ. Both of these proposed cutoffs exhibit unacceptable
dependence on the relative motion of the wormhole mouths.

As an improved alternative cutoff, consider the following: Pick a
point $x$ in spacetime. Since, by hypothesis, the spacetime has
nontrivial topology there will be at least one closed geodesic of
nontrivial homotopy that runs from $x$ to itself. If the length of
this geodesic is less than a Planck length, then the region
surrounding the point $x$ should no longer be treated semiclassically.
Presumably, one should also supplement this requirement by a bound
on the curvature: If $R_{\alpha\beta\gamma\delta}
R^{\alpha\beta\gamma\delta} > \ell_P^{-4}$ then the region surrounding
the point $x$ should no longer be treated semiclassically.

The advantage of the cutoff expressed in this manner is that it
appears to be eminently physically reasonable. This cutoff give
meaningful answers in the case of a relatively stationary pair of
wormhole mouths, does not beg the question by requiring the existence
of a Cauchy horizon to formulate the cutoff, and furthermore can
be concisely and clearly stated in complete generality for arbitrary
spacetimes.

\newpage
\section{DISCUSSION}

This paper has examined at some length a series of physical effects
that lend support to Hawking's Chronology Protection Conjecture.
Explicit calculations have been exhibited for simple models in
suitable parameter regimes.  In particular much has been made of
the use of adiabatic techniques combined with minor variations on
the theme of the Casimir effect.

These calculations suggest that the universe exhibits a defense in
depth strategy with respect to global causality violations. Effects
contributing to the hoped for inability to manufacture a time
machine include vacuum polarization effects, wormhole disruption
effects, and the gravitational back reaction induced by the vacuum
expectation value of the renormalized stress--energy tensor.

Though the calculations in this paper have been phrased in terms
of the traversable wormhole paradigm, the general result to be
abstracted from this analysis is that: (1) any spacetime of nontrivial
topology contains closed (spacelike) geodesics. (2) It appears that
the universe reacts badly to closed spacelike geodesics attempting
to shrink to invariant length zero.

I would conclude by saying that the available evidence now seems
to favour Hawking's Chronology Protection Conjecture.  Both the
experimental evidence (the nonappearance of hordes of tourists from
the future) and the theoretical computations now support this
conjecture. Paraphrasing Hawking, it seems that the universe is
indeed safe for historians.

Perhaps most importantly, the single most serious objection to the
existence of traversable wormholes has always been the {\it apparent}
ease with which they might be converted into time machines. Thus
adopting Hawking's Chronology Protection Conjecture immediately
disposes of the single most serious objection against the existence
of traversable wormholes. This observation should be interpreted
as making the existence of traversable wormholes a much more
reasonable hypothesis. Investigations of questions such as the
Averaged Weak Energy Condition, and all other aspects of traversable
wormhole physics, thus take on a new urgency
\cite{Klinkhammer,Wald-Yurtsever,Yurtsever90a,Yurtsever90b,Yurtsever91}.

\acknowledgements

I wish to thank Carl Bender, Leonid Grischuk, Mike Ogilvie, Donald
Petcher, Tom Roman, and Kip Thorne for useful discussions. I also
wish to thank the Aspen Center for Physics for its hospitality.
This research was supported by the U.S. Department of Energy.


\end{document}